\documentclass[12pt]{article}
\usepackage{latexsym,amssymb,amsfonts,amsmath}
\usepackage{mathrsfs}
\usepackage[dvips]{graphicx}
\newlength{\extraspace}
\newlength{\extraspaces}
\newcommand{\be}{\begin{equation}
\addtolength{\abovedisplayskip}{\extraspaces}
\addtolength{\belowdisplayskip}{\extraspaces}
\addtolength{\abovedisplayshortskip}{\extraspace}
\addtolength{\belowdisplayshortskip}{\extraspace}}
\newcommand{\ee}{\end{equation}}

\newcommand{\bea}{\begin{eqnarray}
\addtolength{\abovedisplayskip}{\extraspaces}
\addtolength{\belowdisplayskip}{\extraspaces}
\addtolength{\abovedisplayshortskip}{\extraspace}
\addtolength{\belowdisplayshortskip}{\extraspace}}
\newcommand{\eea}{\end{eqnarray}}

\DeclareMathOperator{\Tr}{Tr} 
\DeclareMathOperator{\diag}{diag} 

\newcommand{\Overline}[2][1]{%
 {}\mkern#1mu \overline{\mkern-#1mu #2 \mkern-#1mu}\mkern#1mu {}}

\newcommand{\lpq}{\lambda_{\pi}^{2}}
\newcommand{\Lpq}{\Lambda_{\pi}^{2}}

\newcommand{\bpq}{B_{\pi}^{2}}

\begin{document}

\addtolength{\baselineskip}{.8mm}

{\thispagestyle{empty}

\noindent \hspace{1cm} \hfill IFUP--TH/2019 \hspace{1cm}\\

\begin{center}
{\Large\bf Theta dependence of the vacuum energy density}\\
{\Large\bf in chiral effective Lagrangian models}\\
{\Large\bf at finite temperature, above $T_c$}\\
\vspace*{1.0cm}
{\large
Enrico Meggiolaro\footnote{E-mail: enrico.meggiolaro@unipi.it}
}\\
\vspace*{0.5cm}{\normalsize
{Dipartimento di Fisica, Universit\`a di Pisa,
and INFN, Sezione di Pisa,\\
Largo Pontecorvo 3, I-56127 Pisa, Italy}}\\
\vspace*{2cm}{\large \bf Abstract}
\end{center}

\noindent
In this work, extending a previous study at zero temperature ($T=0$),
we perform a systematic study of the modifications to the QCD vacuum energy
density $\epsilon_{vac}$ in the finite-temperature case, above the chiral
transition at $T_c$, caused by a nonzero value of the parameter $\theta$,
using two different effective Lagrangian models which implement the $U(1)$
axial anomaly of the fundamental theory and which are both well defined also
above $T_c$. In particular, we derive (and critically compare) the expressions
for the topological susceptibility $\chi$ and for the second cumulant $c_4$
starting from the $\theta$ dependence of $\epsilon_{vac}(\theta)$
in the two models.

}
\newpage

\section{Introduction}

It is well known (mainly by lattice simulations \cite{HotQCD}) that, at temperatures above a certain critical temperature $T_c \approx 150$ MeV, thermal fluctuations break up the chiral condensate $\langle \bar{q} q \rangle$, causing the complete restoration of the $SU(L)_L\otimes SU(L)_R$ chiral symmetry of QCD with $L$ light quarks ($L=2$ and $L=3$ being the physically relevant cases): this leads to a phase transition called ``chiral transition''.
For what concerns, instead, the $U(1)$ axial symmetry, the nonzero contribution to the anomaly provided by the instanton gas at high temperatures \cite{GPY1981} should imply that it is always broken, also for $T>T_c$. (However, the real magnitude of its breaking and its possible {\it effective} restoration at some temperature above $T_c$ are still important debated questions in hadronic physics.)

In this work, extending a previous study at zero temperature ($T=0$) \cite{LM2018}, we perform a systematic study of the modifications to the QCD vacuum energy density $\epsilon_{vac}$ in the finite-temperature case, above the chiral transition at $T_c$, caused by a nonzero value of the parameter $\theta$, using two different effective Lagrangian models which implement the $U(1)$ axial anomaly of the fundamental theory and which are both well defined also above $T_c$. In particular, we derive (and critically compare) the expressions for the topological susceptibility $\chi$ and for the second cumulant $c_4$ starting from the $\theta$ dependence of $\epsilon_{vac}(\theta)$ in the two models.
Indeed, these two quantities are known to be, respectively, the second and the fourth derivative with respect to $\theta$ of the vacuum energy density, evaluated at $\theta=0$: $\epsilon_{vac}(\theta)=\,const. + \frac{1}{2}\chi\theta^2 + \frac{1}{24}c_4\theta^4 + \ldots$.


The first effective Lagrangian model that we shall consider was originally proposed in Ref. \cite{ELSM1} to study the chiral dynamics at $T=0$, and later used as an effective model to study the chiral-symmetry restoration at nonzero temperature \cite{PW1984,ELSMfiniteT_1,ELSMfiniteT_2}.
According to 't Hooft (see Refs. \cite{ELSM2,ELSM3} and references therein), it reproduces, in terms of an effective theory, the $U(1)$ axial breaking caused by instantons in the fundamental theory.\footnote{We recall here, however, the criticism by Christos \cite{Christos1984} (see also Refs. \cite{WDV1,WDV2}), according to which the determinantal interaction term in this effective model [see Eq. \eqref{Finite temperature ELsm: potential L>2} below] does not correctly reproduce the $U(1)$ axial anomaly of the fundamental theory.}
For brevity, following the notation already introduced in Ref. \cite{LM2018},
we shall refer to it as the ``extended linear sigma ($EL_\sigma$) model''. This model is described by the following Lagrangian:
\begin{equation}\label{'t Hooft effective Lagrangian}
\mathscr{L}_{(EL_\sigma)}(U,U^{\dagger}) = \frac{1}{2}\Tr [\partial_\mu U \partial^\mu U^{\dagger}] - V(U,U^{\dagger}) ,
\end{equation}
where
\begin{equation}\label{Finite temperature ELsm: potential L>2}
\begin{aligned}
V(U,U^\dagger) &= \frac{1}{4}\lambda_\pi^2 \Tr \left[(UU^\dagger - \rho_\pi \mathbf{I})^2\right] + \frac{1}{4}\lambda_\pi^{'2} \left[\Tr(UU^\dagger)\right]^2 \\ &- \frac{B_m}{2\sqrt{2}}\Tr \left[\mathcal{M}U + \mathcal{M}^\dagger U^\dagger\right] - \kappa \left[\det U + \det U^\dagger\right] .
\end{aligned}
\end{equation}
In this model, the mesonic effective fields are represented by a $L\times L$ complex matrix $U_{ij}$ which can be written in terms of the quark fields as $U_{ij}\sim \Overline[2]{q}_{jR}q_{iL}$, up to a multiplicative constant; moreover, $\mathcal{M}$ is a complex quark-mass matrix, given by
\begin{equation}\label{mass matrix with theta term}
\mathcal{M}=Me^{i\frac{\theta}{L}} ,
\end{equation}
where $M = \diag (m_1, \ldots, m_L)$ is the physical (real and diagonal) quark-mass matrix.
(In this paper, therefore, we have decided to move all the dependence on $\theta$ into the mass term, for later convenience.)

For what concerns the potential $V(U,U^\dagger)$ defined in Eq. \eqref{Finite temperature ELsm: potential L>2}, we recall that the parameter $\rho_\pi$ is responsible for the fate of the $SU(L)_L \otimes SU(L)_R$ chiral symmetry, which, as is well known, depends on the temperature $T$.
We shall include the effects of the temperature in the model allowing the various parameters in Eq. \eqref{Finite temperature ELsm: potential L>2} to vary with the temperature: in particular, the parameter $\rho_\pi$ will be positive, and, correspondingly, the ``vacuum expectation value'' (vev), i.e., the thermal average, of $U$ will be different from zero in the chiral limit $M=0$, until the temperature reaches the chiral phase-transition temperature $T_c$ [$\rho_\pi(T<T_c)>0$], above which it will be negative [$\rho_\pi(T>T_c)<0$], and, correspondingly, the vev of $U$ will vanish in the chiral limit $M=0$.\footnote{We notice here that we have identified the temperature $T_{\rho_\pi}$ at which the parameter $\rho_\pi$ is equal to zero with the chiral phase-transition temperature $T_c$: this is always correct except in the case $L=2$, where we have $T_{\rho_\pi} < T_c$ (see Secs. 2.2 and 3.2 for a more detailed discussion). In any case, in this paper we shall consider exclusively the region of temperatures $T > T_c$.}


The second effective Lagrangian model that we shall consider is a generalization of the model proposed by Witten, Di Vecchia, Veneziano, \emph{et al.} \cite{WDV1,WDV2,WDV3} (that, following the notation introduced in Ref. \cite{LM2018}, will be denoted for brevity as the ``WDV model''), and (in a sense which will be made clear below) it approximately ``interpolates`` between the WDV model at $T=0$ and the $EL_\sigma$ model for $T>T_c$: for this reason (always following Ref. \cite{LM2018}) we shall call it the ``interpolating model'' (IM).
In this model (which was originally proposed in Ref. \cite{EM1994} and elaborated on in Refs. \cite{MM2003,EM2011,MM2013}), the $U(1)$ axial anomaly is implemented, as in the WDV model, by properly introducing the topological charge density $Q(x)=\frac{g^{2}}{64\pi^{2}}\varepsilon^{\mu\nu\rho\sigma} F_{\mu\nu}^{a}(x)F_{\rho\sigma}^{a}(x)$ as an auxiliary field, so that it satisfies the correct transformation property under the chiral group.\footnote{However, we must recall here that also the particular way of implementing the $U(1)$ axial anomaly in the WDV model, by means of a logarithmic interaction term [as in Eqs. \eqref{Interpolating model Lagrangian with Q} and \eqref{potential of the interpolating model after having integrated out Q} below], was criticized by 't Hooft in Ref. \cite{ELSM2}. Unfortunately, no real progress has been done up to now to solve the controversy (recalled also in the first footnote) between Ref. \cite{ELSM2} and Ref. \cite{Christos1984}, and we are still living with it.}
Moreover, it also assumes that there is another $U(1)$-axial-breaking condensate (in addition to the usual quark-antiquark chiral condensate $\langle \bar{q}q \rangle$), having the form $C_{U(1)} = \langle {\cal O}_{U(1)} \rangle$,
where, for a theory with $L$ light quark flavors, ${\cal O}_{U(1)}$ is a
$2L$-quark local operator that has the chiral transformation properties of
\cite{tHooft1976,KM1970,Kunihiro2009}
${\cal O}_{U(1)} \sim \displaystyle{{\det_{st}}(\bar{q}_{sR}q_{tL})
+ {\det_{st}}(\bar{q}_{sL}q_{tR}) }$,
where $s,t = 1, \ldots, L$ are flavor indices.\footnote{The color indices (not
explicitly indicated) are arranged in such a way that
(i) ${\cal O}_{U(1)}$ is a color singlet, and (ii)
$C_{U(1)} = \langle {\cal O}_{U(1)} \rangle$ is a \emph{genuine} $2L$-quark
condensate, i.e., it has no \emph{disconnected} part proportional to some
power of the quark-antiquark chiral condensate $\langle \bar{q} q \rangle$;
the explicit form of the condensate for the cases $L=2$ and $L=3$ is
discussed in detail in the Appendix A of Ref. \cite{EM2011}.}

The effective Lagrangian of the interpolating model is written in terms
of the topological charge density $Q$, the mesonic field
$U_{ij} \sim \bar{q}_{jR} q_{iL}$ (up to a multiplicative constant),
and the new field variable $X \sim {\det} \left( \bar{q}_{sR} q_{tL} \right)$
(up to a multiplicative constant), associated with the $U(1)$ axial
condensate:
\begin{equation}\label{Interpolating model Lagrangian with Q}
\begin{split}
\mathscr{L}_{(IM)}(U,&U^\dagger,X,X^\dagger,Q)
=\frac{1}{2}\Tr [\partial_\mu U \partial^\mu U^{\dagger}]
+ \frac{1}{2}\partial_\mu X \partial^\mu X^{\dagger}
-V_0(U,U^\dagger,X,X^\dagger) \\
&+ \frac{i}{2}Q \left[ \omega_1 \Tr (\log U -\log U^\dagger)
+ (1-\omega_1) (\log X -\log X^\dagger) \right] + \frac{1}{2A}Q^2 ,
\end{split}
\end{equation}
where
\begin{equation}\label{potential of the interpolating model}
\begin{split}
V_0(U,U^\dagger,X,X^\dagger) &= \frac{1}{4}\lambda_\pi^2\Tr [(UU^{\dagger}-\rho_\pi \mathbf{I})^2] + \frac{1}{4}\lambda_\pi^{'2}\left[\Tr(UU^\dagger)\right]^2 + \frac{1}{4}\lambda_X^2 [XX^\dagger - \rho_X]^2 \\ &- \frac{B_m}{2\sqrt{2}}\Tr\left[\mathcal{M} U + \mathcal{M}^\dagger U^{\dagger}\right] - \frac{\kappa_1}{2\sqrt{2}}[X^\dagger \det U + X\det U^\dagger ] . 
\end{split}
\end{equation}
Once again, we have decided (for later convenience) to put all the $\theta$ dependence in the complex mass matrix $\mathcal{M}=Me^{i\frac{\theta}{L}}$.\\
As in the case of the WDV model, the auxiliary field $Q$ in \eqref{Interpolating model Lagrangian with Q} can be integrated out using its equation of motion:
\begin{equation}
Q = -\frac{i}{2} A \left[\omega_1 \Tr (\log U - \log U^\dagger )+ (1-\omega_1)(\log X - \log X^\dagger) \right] .
\end{equation}
After the substitution, we obtain
\begin{equation}\label{Interpolating model Lagrangian without Q}
\mathscr{L}_{(IM)}(U,U^\dagger,X,X^\dagger)=\frac{1}{2}\Tr [\partial_\mu U \partial^\mu U^{\dagger}] + \frac{1}{2}\partial_\mu X \partial^\mu X^{\dagger}
-V(U,U^\dagger , X,X^\dagger) ,
\end{equation}
where
\begin{equation}\label{potential of the interpolating model after having integrated out Q}
\begin{split}
V(U,&U^\dagger,X,X^\dagger)=V_0(U,U^\dagger,X,X^\dagger) \\
&-\frac{1}{8} A \left[\omega_1 \Tr (\log U - \log U^\dagger)+ (1-\omega_1)(\log X - \log X^\dagger )\right]^2 .
\end{split}
\end{equation}
All the parameters which appear in Eqs.
\eqref{potential of the interpolating model} and
\eqref{potential of the interpolating model after having integrated out Q}
have to be considered as temperature dependent. In particular, the parameter $\rho_X$ plays for the $U(1)$ axial symmetry the same role the parameter $\rho_\pi$ plays for the $SU(L)_L \otimes SU(L)_R$ chiral symmetry: $\rho_X$ determines the vev of the field $X$ and it is thus responsible for the way in which the $U(1)$ axial symmetry is realised. In order to reproduce the scenario we are interested in, that is, the scenario in which the $U(1)$ axial symmetry is {\it not} restored for $T>T_c$, while the $SU(L) \otimes SU(L)$ chiral symmetry is restored as soon as the temperature reaches $T_c$, we must assume that, differently from $\rho_\pi$, the parameter $\rho_X$ remains positive across $T_c$, i.e., $\rho_\pi(T<T_c)>0$, $\rho_X(T<T_c)>0$, and $\rho_\pi(T>T_c)<0$, $\rho_X(T>T_c)>0$.

For what concerns the parameter $\omega_1(T)$, in order to avoid a singular behavior of the anomalous term in the potential
\eqref{potential of the interpolating model after having integrated out Q}
above the chiral-transition temperature $T_c$, where the vev of the mesonic field $U$ vanishes (in the chiral limit $M=0$), we must assume that \cite{EM1994,MM2013} $\omega_1 (T\geq T_c)=0$.\\
(This way, indeed, the term including $\log U$ in the potential vanishes, eliminating the problem of the divergence, at least as far as the vev of the field $X$ is different from zero or, in other words, as far as the $U(1)$ axial symmetry remains broken also above $T_c$.)

As it was already observed in Refs. \cite{EM2011,LM2018}, the Lagrangian of the WDV model is obtained from that of the interpolating model by first fixing $\omega_1=1$ and then taking the formal limits $\lambda_X \to +\infty$ and also $\rho_X \to 0$ (so that $X \to 0$):
\begin{equation}\label{IM to WDV}
\mathscr{L}_{(IM)}\vert_{\omega_1=1} \mathop{\longrightarrow}_{\lambda_X \to +\infty,~\rho_X \to 0} \mathscr{L}_{(WDV)} .
\end{equation}
For this reason, $\omega_1=1$ seems to be the most natural choice for $T=0$ (and, indeed, it was found in Ref. \cite{LM2018} that the expressions for $\chi$ and $c_4$, obtained using the interpolating model with $\omega_1=1$, coincide with those of the WDV model, \emph{regardless} of the values of the other parameters $\kappa_1$ and $\rho_X$\dots).\\
On the other side, as we have seen above, the parameter $\omega_1$ must be necessarily taken to be equal to zero above the critical temperature $T_c$, where the WDV is no more valid (because of the singular behavior of the anomalous term in the potential), and vice versa, as it was already observed in Ref. \cite{MM2013}, the interaction term $\frac{\kappa_1}{2\sqrt{2}}[X^\dagger \det U + X\det U^\dagger]$ of the interpolating model becomes very similar to the ``instantonic'' interaction term $\kappa [\det U + \det U^\dagger]$ of the $EL_\sigma$ model.
More precisely, we here observe that, by first fixing $\omega_1=0$ and then taking the formal limits $\lambda_X \to +\infty$ and $A \to \infty$ (so that, writing $X= \alpha e^{i\beta}$, one has $\alpha \to \sqrt{\rho_X}$ and $\beta \to 0$, i.e., $X \to \sqrt{\rho_X}$), the Lagrangian of the interpolating model reduces to the Lagrangian of the $EL_\sigma$ model with $\kappa=\frac{\kappa_1\sqrt{\rho_X}}{2\sqrt{2}}$ (i.e., with $\kappa$ proportional to the $U(1)$ axial condensate):
\begin{equation}\label{IM to ELsm}
\mathscr{L}_{(IM)}\vert_{\omega_1=0} \mathop{\longrightarrow}_{\lambda_X \to +\infty,~A \to +\infty} \mathscr{L}_{(EL_\sigma)}\vert_{\kappa=\frac{\kappa_1\sqrt{\rho_X}}{2\sqrt{2}}} .
\end{equation}
The paper is organized as follows: in Secs. 2 and 3 we shall present the results for the extended linear sigma model and the interpolating model, respectively. These results will be obtained at the first nontrivial order in an expansion in the quark masses (since this will greatly simplify the search for the minimum of the potential). On the other side, no assumption will be done on the parameter $\theta$, which will be treated as an absolutely free parameter.
Moreover, for each of the two models considered, we shall present separately the results for the cases $L\geq 3$ and $L=2$, due to the fact that (for some technical reasons which will be explained in the following: see also Ref. \cite{MM2013}) the case $L=2$ requires a more specific analysis.
Finally, in the last section we shall draw our conclusions, summarizing (and critically commenting) the results obtained in this work and discussing also some possible future developments.

\section{Results for the extended linear sigma model}

\subsection{The case $L\geq 3$}

Following the notation of Ref. \cite{MM2013}, we shall write the parameter $\rho_\pi$, for $T>T_c$, as follows:
\begin{equation}\label{rho_pi for T>Tc}
\rho_\pi \equiv -\frac{1}{2}B_\pi^2<0 ,
\end{equation}
and, moreover, we shall use for the matrix field $U$ the following simple linear parametrization:
\begin{equation}\label{linear parametrization}
U_{ij} = a_{ij} + ib_{ij} ,
\end{equation}
where $a_{ij}$ and $b_{ij}$ are real field variables whose vevs $\bar{a}_{ij}$ and $\bar{b}_{ij}$ vanish in the chiral limit ($\Overline[2]{U}=0$ for $M=0$, when $T>T_c$).
We shall also write the complex mass matrix \eqref{mass matrix with theta term}
in a similar way, i.e., separating its real and imaginary parts:
\begin{equation}\label{Mass matrix form for T>Tc}
\mathcal{M}_{ij} = M_{ij}\, e^{i\frac{\theta}{L}} \equiv m_{ij} + i n_{ij} .
\end{equation}
With this choice of the parametrizations for the parameter $\rho_\pi$, the fields $U$, and the mass matrix $\mathcal{M}$, the potential \eqref{Finite temperature ELsm: potential L>2} becomes
\begin{equation}\label{Finite temperature ELsm: explicit potential L>2}
\begin{aligned}
V &= \frac{L}{16}\lambda_\pi^2 B_\pi^4 + \frac{1}{4}\lambda_\pi^2 B_\pi^2(a_{ij}^2+b_{ij}^2) -\frac{B_m}{\sqrt{2}}(m_{ij}a_{ji} - n_{ij}b_{ji}) \\
&+ \frac{1}{4}\lambda_\pi^2 \Tr \left[(UU^\dagger)^2\right] + \frac{1}{4}\lambda_\pi^{'2} \left[\Tr(UU^\dagger)\right]^2 - \kappa \left[\det U + \det U^\dagger\right] .
\end{aligned}
\end{equation}
In order to find the value $\Overline[2]{U}$ for which the potential $V$ is minimum (that is, in our mean-field approach, the vev of $U$), we have to solve the following system of stationary-point equations:
\begin{equation}\label{Finite temperature ELsm: minimization system L>2}
\left\{
\begin{aligned}
\left.\frac{\partial V}{\partial a_{ij}}\right|_S &= \frac{1}{2}\lambda_\pi^2 B_\pi^2 \,\bar{a}_{ij} - \frac{B_m}{\sqrt{2}}m_{ji} + \ldots = 0 ,\\
\left.\frac{\partial V}{\partial b_{ij}}\right|_S &= \frac{1}{2}\lambda_\pi^2 B_\pi^2 \,\bar{b}_{ij} + \frac{B_m}{\sqrt{2}}\,n_{ji} + \ldots = 0 ,
\end{aligned}
\right.
\end{equation}
where the neglected terms are of quadratic or higher order in the fields. We can easily solve this system, at the leading order in the quark masses, obtaining:
$\Overline[2]{U}_{ij} = \bar{a}_{ij} + i \bar{b}_{ij} \simeq \frac{2B_m}{\sqrt{2}\lambda_\pi^2 B_\pi^2}\, (m_{ji}-in_{ji})$, that is
\begin{equation}\label{Vev of U L>2}
\Overline[2]{U} \simeq \frac{2B_m}{\sqrt{2}\lambda_\pi^2 B_\pi^2} \mathcal{M}^\dagger = \frac{2B_m}{\sqrt{2}\lambda_\pi^2 B_\pi^2} M e^{-i\frac{\theta}{L}} .
\end{equation}
A simple analysis of the second derivatives of the potential $V$ with respect to the fields, calculated in this point, confirms that it is indeed a minimum of the potential.
So, we find that (at the first nontrivial order in the quark masses) the vev of the mesonic field $U$ is proportional to the mass matrix. We notice here that, by virtue of the result \eqref{Vev of U L>2}, the quantities $\Overline[2]{U}\Overline[2]{U}^\dagger$ and $\mathcal{M}\Overline[2]{U}$ turn out to be independent of $\theta$. Therefore, all the terms of the potential \eqref{Finite temperature ELsm: potential L>2} carry no dependence on $\theta$ except for the ``instantonic'' one. That is, explicitly,
\begin{equation}\label{Finite temperature ELsm: theta dependence of potential L>2}
\begin{aligned}
V_{min}(\theta) &= V(\Overline[2]{U}(\theta)) = const. -\kappa \left(\det \Overline[2]{U}(\theta) + \det \Overline[2]{U}^\dagger(\theta)\right) + \ldots \\
&= const. - 2\kappa\left(\frac{2B_m}{\sqrt{2}\lambda_\pi^2 B_\pi^2}\right)^L \det M \cos\theta + \ldots ,
\end{aligned}
\end{equation}
where the omitted terms are either constant with respect to $\theta$ or of higher order in the quark masses. Finally, from \eqref{Finite temperature ELsm: theta dependence of potential L>2} we can straightforwardly derive the topological susceptibility and the second cumulant, which turn out to be
\begin{equation}\label{Finite temperature ELsm: chi and c4 L>2}
\begin{aligned}
\chi = \left.\frac{\partial^2 V_{min}(\theta)}{\partial \theta^2}\right|_{\theta=0} &\simeq 2\kappa\left(\frac{2B_m}{\sqrt{2}\lambda_\pi^2 B_\pi^2}\right)^L \det M ,\\
c_4 = \left.\frac{\partial^4 V_{min}(\theta)}{\partial \theta^4}\right|_{\theta=0} &\simeq -2\kappa\left(\frac{2B_m}{\sqrt{2}\lambda_\pi^2 B_\pi^2}\right)^L \det M .
\end{aligned}
\end{equation}


\subsection{The special case $L=2$}

As already said in the Introduction, the case $L=2$ requires a more specific analysis. In fact, in this case, the determinant of the matrix field $U$ is quadratic in the fields and so it must be considered explicitly in the stationary-point equations at the leading order in the quark masses.
In this particular case, it is more convenient to choose for the parametrization of the field $U$ a variant of the linear parametrization \eqref{linear parametrization}, which is explicitly written in terms of the fields describing the mesonic excitations $\sigma$, $\eta$, $\vec{\delta}$ and $\vec{\pi}$, i.e.,
\begin{equation}\label{linear parametrization for L=2}
U=\frac{1}{\sqrt{2}}\left[(\sigma + i \eta )\mathbf{I}+ (\vec{\delta}+i\vec{\pi})\cdot \vec{\tau}\right] ,
\end{equation}
where $\tau_a$ ($a=1,2,3$) are the Pauli matrices (with the usual normalization $\Tr [\tau_a\tau_b] = 2\delta_{ab}$), while the multiplicative factor $\frac{1}{\sqrt{2}}$ guarantees the correct normalization of the kinetic term in the effective Lagrangian.
We expect that all the vevs of the fields $\sigma$, $\eta$, $\vec{\delta}$ and $\vec{\pi}$ are (at the leading order) proportional to the quark masses, so that they vanish in the chiral limit $M\to 0$.
Using the parametrization \eqref{linear parametrization for L=2}, we find the following expression
for the potential \eqref{Finite temperature ELsm: potential L>2} (having defined $\Lambda_\pi^2 \equiv \lambda_\pi^2 + 2 \lambda_\pi^{'2}$):
\begin{equation}\label{Finite temperature ELsm: explicit potential L=2}
\begin{aligned}
V &= \frac{1}{8}\lambda_\pi^2 B_\pi^4 + \frac{1}{8}\Lambda_\pi^2(\sigma^2 + \eta^2 + \vec{\delta}^2 + \vec{\pi}^2)^2 + \frac{1}{2}\lambda_\pi^2(\sigma^2\vec{\delta}^2 {+} 2\sigma\eta\vec{\delta}\cdot\vec{\pi} {+} \eta^2\vec{\pi}^2) \\
&+ \frac{1}{2}\lambda_\pi^2\left[\vec{\pi}^2 \vec{\delta}^2 {-} (\vec{\delta}\cdot\vec{\pi})^2\right] + \frac{1}{4}\lambda_\pi^2 B_\pi^2(\sigma^2 + \eta^2 + \vec{\delta}^2 + \vec{\pi}^2) \\
&- \frac{B_m}{2}\left[(m_u {+} m_d)\left(\sigma\cos\frac{\theta}{2} {-} \eta\sin\frac{\theta}{2}\right) {+} (m_u{-}m_d)\left(\delta_3\cos\frac{\theta}{2} {-} \pi_3\sin\frac{\theta}{2}\right)\right] \\
&- \kappa(\sigma^2 - \eta^2 - \vec{\delta}^2 + \vec{\pi}^2) .
\end{aligned}
\end{equation}
We now look for the minimum of the potential, solving the following system of stationary-point equations:
\begin{equation}\label{Finite temperature ELsm: minimization system L=2}
\left\{
\begin{aligned}
& \left.\frac{\partial V}{\partial \sigma}\right|_S = \frac{1}{2}\Lambda_\pi^2\left(\bar{\sigma}^2 + \bar{\eta}^2 + \bar{\vec{\delta}}^2 + \bar{\vec{\pi}}^2\right)\bar{\sigma} + \lambda_\pi^2\left(\bar{\sigma}\bar{\vec{\delta}}^2 + \bar{\eta}\bar{\vec{\delta}}\cdot\bar{\vec{\pi}}\right) \\
& \hspace{2 cm} + \frac{1}{2}(\lambda_\pi^2 B_\pi^2 - 4\kappa)\bar{\sigma}
-\frac{B_m}{2}(m_u + m_d)\cos\frac{\theta}{2} = 0 ,\\ \\
& \left.\frac{\partial V}{\partial \eta}\right|_S = \frac{1}{2}\Lambda_\pi^2\left(\bar{\sigma}^2 + \bar{\eta}^2 + \bar{\vec{\delta}}^2 + \bar{\vec{\pi}}^2\right)\bar{\eta} + \lambda_\pi^2\left(\bar{\sigma}\bar{\vec{\delta}}\cdot\bar{\vec{\pi}} + \bar{\eta}\bar{\vec{\pi}}^2\right) \\
& \hspace{2 cm} + \frac{1}{2}(\lambda_\pi^2 B_\pi^2 + 4\kappa)\bar{\eta}
+\frac{B_m}{2}(m_u + m_d)\sin\frac{\theta}{2} = 0 ,\\ \\
& \left.\frac{\partial V}{\partial \delta_a}\right|_S = \frac{1}{2}\Lambda_\pi^2\left(\bar{\sigma}^2 {+} \bar{\eta}^2 {+} \bar{\vec{\delta}}^2 {+} \bar{\vec{\pi}}^2\right)\bar{\delta}_a {+} \lambda_\pi^2\left(\bar{\sigma}^2\bar{\delta}_a {+} \bar{\sigma}\bar{\eta}\bar{\pi}_a\right) {+} \lambda_\pi^2\left[\bar{\vec{\pi}}^2\bar{\delta}_a {-} (\bar{\vec{\pi}}\cdot\bar{\vec{\delta}})\bar{\pi}_a\right] \\ & \hspace{2 cm} +\frac{1}{2}(\lambda_\pi^2 B_\pi^2 + 4\kappa)\bar{\delta}_a-\frac{B_m}{2}(m_u - m_d)\cos\frac{\theta}{2}\delta_{a3} = 0 ,\\ \\
& \left.\frac{\partial V}{\partial \pi_a}\right|_S = \frac{1}{2}\Lambda_\pi^2\left(\bar{\sigma}^2 {+} \bar{\eta}^2 {+} \bar{\vec{\delta}}^2 {+} \bar{\vec{\pi}}^2\right)\bar{\pi}_a {+} \lambda_\pi^2\left(\bar{\sigma}\bar{\eta}\bar{\delta}_a{+}\bar{\eta}^2\bar{\pi}_a\right) {+} \lambda_\pi^2\left[\bar{\vec{\delta}}^2\bar{\pi}_a {-} (\bar{\vec{\pi}}\cdot\bar{\vec{\delta}})\bar{\delta}_a\right] \\ & \hspace{2 cm} + \frac{1}{2}(\lambda_\pi^2 B_\pi^2 - 4\kappa)\bar{\pi}_a+\frac{B_m}{2}(m_u - m_d)\sin\frac{\theta}{2}\delta_{a3} = 0 .
\end{aligned}
\right.
\end{equation}
Solving these equations at the first nontrivial order in the quark masses, one immediately finds that $\bar{\delta}_1 = \bar{\delta}_2 = \bar{\pi}_1 = \bar{\pi}_2 =0$ (i.e., the matrix field $\Overline[2]{U}$ turns out to be diagonal, as expected, being the mass matrix $\mathcal{M} = M e^{i\frac{\theta}{2}}$ diagonal), and moreover
\begin{equation}\label{Finite temperature ELsm: solution of the fields L=2}
\begin{aligned}
\bar{\sigma} &\simeq \frac{B_m (m_u+m_d)}{\lambda_\pi^2 B_\pi^2 - 4 \kappa}\cos\frac{\theta}{2} ,\qquad \bar{\eta} \simeq -\frac{B_m (m_u+m_d)}{\lambda_\pi^2 B_\pi^2 + 4 \kappa}\sin\frac{\theta}{2} ,\\
\bar{\delta}_3 &\simeq \frac{B_m (m_u-m_d)}{\lambda_\pi^2 B_\pi^2 + 4 \kappa}\cos\frac{\theta}{2} ,\qquad \bar{\pi}_3 \simeq -\frac{B_m (m_u-m_d)}{\lambda_\pi^2 B_\pi^2 - 4 \kappa}\sin\frac{\theta}{2} .
\end{aligned}
\end{equation}
Studying the matrix of the second derivatives of the potential with respect to the fields, one immediately sees that this stationary point corresponds indeed to a minimum of the potential, provided that the condition $\lambda_\pi^2 B_\pi^2 > 4\kappa$ is satisfied.
Remembering Eq. \eqref{rho_pi for T>Tc}, this condition can be written as ${\cal G}_\pi \equiv 4\kappa + 2\lambda_\pi^2 \rho_\pi < 0$ and the ``critical transition temperature'' $T_c$ is just defined by the condition ${\cal G}_\pi(T=T_c)=0$: assuming that $\kappa>0$, this implies that in this case (differently from the case $L \ge 3$) $T_c>T_{\rho_\pi}$, where $T_{\rho_\pi}$ is defined to be the temperature at which $\rho_\pi$ vanishes [with $\rho_\pi(T<T_{\rho_\pi})>0$ and $\rho_\pi(T>T_{\rho_\pi})<0$; see also Ref. \cite{MM2013} for a more detailed discussion on this question].

Substituting the solution \eqref{Finite temperature ELsm: solution of the fields L=2} into Eq. \eqref{Finite temperature ELsm: explicit potential L=2} (and neglecting, for consistency, all the terms which are more than quadratic in the quark masses or which are simply constant with respect to $\theta$), we find the following $\theta$ dependence for the minimum value of the potential:
\begin{equation}\label{Finite temperature ELsm: theta dependence of potential L=2}
\begin{aligned}
V_{min}(\theta) &= \frac{1}{8}\lambda_\pi^2 B_\pi^4 {+} \frac{1}{4}\lambda_\pi^2 B_\pi^2(\bar{\sigma}^2 {+} \bar{\eta}^2 {+} \bar{\delta}_3^2 {+} \bar{\pi}_3^2) {-}\frac{B_m}{2}\left[(m_u{+}m_d)\left(\bar{\sigma}\cos\frac{\theta}{2}{-} \bar{\eta}\sin\frac{\theta}{2}\right) \right. \\
&+ \left.(m_u {-} m_d)\left(\bar{\delta}_3\cos\frac{\theta}{2}{-} \bar{\pi}_3\sin\frac{\theta}{2}\right)\right] {-}\kappa(\bar{\sigma}^2 {-} \bar{\eta}^2 {-} \bar{\delta}_3^2 {+} \bar{\pi}_3^2) + O(m^3) \\
&= \, const. -\frac{4\kappa B_m^2 m_u m_d}{\lambda_\pi^4 B_\pi^4 - 16\kappa^2}\cos\theta + O(m^3) .
\end{aligned}
\end{equation}
From Eq. \eqref{Finite temperature ELsm: theta dependence of potential L=2}, we can derive the following expressions of the topological susceptibility and of the second cumulant:
\begin{equation}\label{Finite temperature ELsm: chi and c4 L=2}
\begin{aligned}
\chi = \left.\frac{\partial^2 V_{min}(\theta)}{\partial \theta^2}\right|_{\theta=0} &\simeq \frac{4\kappa B_m^2}{\lambda_\pi^4 B_\pi^4 - 16\kappa^2} m_u m_d ,\\
c_4 = \left.\frac{\partial^4 V_{min}(\theta)}{\partial \theta^4}\right|_{\theta=0} &\simeq -\frac{4\kappa B_m^2}{\lambda_\pi^4 B_\pi^4 - 16\kappa^2} m_u m_d .
\end{aligned}
\end{equation}

\section{Results for the interpolating model with the inclusion of a $U(1)$ axial condensate}

\subsection{The case $L\geq 3$}

Following, as usual, the notation of Ref. \cite{MM2013}, we shall write the parameters $\rho_\pi$ and $\rho_X$ for $T>T_c$ as follows:
\begin{equation}\label{rho_pi and rho_X for T>Tc}
\rho_\pi \equiv -\frac{1}{2}B_\pi^2<0 ,~~~ \rho_X \equiv \frac{1}{2}F_X^2>0 .
\end{equation}
Moreover, we shall continue to write the complex mass matrix in the
form \eqref{Mass matrix form for T>Tc} and, concerning the fields, we shall
use for $U$ the usual linear parametrization \eqref{linear parametrization},
while we shall use for $X$ the following nonlinear parametrization (in the
form of a {\it polar decomposition}):
\begin{equation}\label{Parametrization of the field X}
X = \alpha e^{i\beta} .
\end{equation}
With this choice of the parametrizations for the parameters $\rho_\pi$ and $\rho_X$, the fields $U$ and $X$, and the mass matrix $\mathcal{M}$, the following expression for the potential of the interpolating model at $T>T_c$ is found:
\begin{equation}\label{Finite temperature Interpolating: explicit potential L>2}
\begin{aligned}
V &= \frac{L}{16}\lambda_\pi^2 B_\pi^4 + \frac{1}{4}\lambda_\pi^2 B_\pi^2(a_{ij}^2+b_{ij}^2) + \frac{1}{4}\lambda_X^2\left(\alpha^2 - \frac{1}{2}F_X^2\right)^2 + \frac{1}{2}A\beta^2 \\
&-\frac{B_m}{\sqrt{2}}(m_{ij}a_{ji} - n_{ij}b_{ji})+ \frac{1}{4}\lambda_\pi^2 \Tr \left[(UU^\dagger)^2\right] + \frac{1}{4}\lambda_\pi^{'2} \left[\Tr(UU^\dagger)\right]^2 \\
&-\frac{\kappa_1\alpha}{2\sqrt{2}} \left[\cos\beta(\det U + \det U^\dagger)-i\sin\beta(\det U - \det U^\dagger)\right] .
\end{aligned}
\end{equation}
The minimum of the potential is found by solving the following system of stationary-point equations:
\begin{equation}\label{Finite temperature Interpolating: minimization system L>2}
\left\{
\begin{aligned}
\left.\frac{\partial V}{\partial a_{ij}}\right|_S &= \frac{1}{2}\lambda_\pi^2 B_\pi^2 \,\bar{a}_{ij} - \frac{B_m}{\sqrt{2}}m_{ji} + \ldots = 0 ,\\ \\
\left.\frac{\partial V}{\partial b_{ij}}\right|_S &= \frac{1}{2}\lambda_\pi^2 B_\pi^2 \,\bar{b}_{ij} + \frac{B_m}{\sqrt{2}}\,n_{ji} + \ldots = 0 ,\\ \\
\left.\frac{\partial V}{\partial \alpha}\right|_S &= \lambda_X^2 \left(\bar{\alpha}^2 - \frac{F_X^2}{2}\right)\bar{\alpha} \\
&- \frac{\kappa_1}{2\sqrt{2}} \left[\cos\bar{\beta}(\det \Overline[2]{U} + \det \Overline[2]{U}^\dagger)-i\sin\bar{\beta}(\det \Overline[2]{U} - \det \Overline[2]{U}^\dagger)\right] = 0 ,\\ \\
\left.\frac{\partial V}{\partial \beta}\right|_S &= A \bar{\beta}
+ \frac{\kappa_1\bar{\alpha}}{2\sqrt{2}} \left[\sin\bar{\beta}(\det \Overline[2]{U} + \det \Overline[2]{U}^\dagger)+i\cos\bar{\beta}(\det \Overline[2]{U} - \det \Overline[2]{U}^\dagger)\right] = 0 .
\end{aligned}
\right.
\end{equation}
We notice that the first two equations \eqref{Finite temperature Interpolating: minimization system L>2} coincide with the equations \eqref{Finite temperature ELsm: minimization system L>2}, so that the solution for $\bar{a}_{ij}$ and $\bar{b}_{ij}$ (i.e., for $\Overline[2]{U}$) will be, at the leading order in the quark masses, exactly the same that has been found in the $EL_\sigma$ model [see Eq. \eqref{Vev of U L>2}].
Moreover, with that expression for $\Overline[2]{U}$, we can see that $\det \Overline[2]{U} + \det \Overline[2]{U}^\dagger \sim \det M \cos\theta \quad$ and $\det \Overline[2]{U} - \det \Overline[2]{U}^\dagger \sim \det M \sin\theta$,
and from the second couple of equations \eqref{Finite temperature Interpolating: minimization system L>2} we can conclude that $\bar{\alpha} \sim \frac{F_X}{\sqrt{2}} + O(\det M \cos\theta)$ and $\bar{\beta} \sim O(\det M \sin\theta)$. More precisely, we find that\footnote{Studying the matrix of the second derivatives, one easily sees that the solution \eqref{Finite temperature Interpolating: alpha beta L>2} for $\bar{\alpha}$ and $\bar{\beta}$ (which, in the chiral limit, reduces to $\bar{\alpha} = \frac{F_X}{\sqrt{2}}$ and $\bar{\beta}=0$) indeed corresponds to the minimum of the potential (see also Ref. \cite{MM2013} for more details).}
\begin{equation}\label{Finite temperature Interpolating: alpha beta L>2}
\begin{aligned}
\bar{\alpha} &\simeq \frac{F_X}{\sqrt{2}} + \frac{\kappa_1}{\sqrt{2}\lambda_X^2 F_X^2} \left(\frac{2B_m}{\sqrt{2}\lambda_\pi^2 B_\pi^2}\right)^L \det M \cos\theta ,\\
\bar{\beta} &\simeq -\frac{1}{A}\frac{\kappa_1F_X}{2} \left(\frac{2B_m}{\sqrt{2}\lambda_\pi^2 B_\pi^2}\right)^L \det M \sin\theta .
\end{aligned}
\end{equation}
We can now substitute the solutions \eqref{Vev of U L>2} and \eqref{Finite temperature Interpolating: alpha beta L>2} into the expression \eqref{Finite temperature Interpolating: explicit potential L>2}, in order to find the $\theta$ dependence of the minimum value of the potential.
As in the case of the $EL_\sigma$ model, the mass term and the terms dependent only on the quantity $\Overline[2]{U}\Overline[2]{U}^\dagger$ turn out to be independent of $\theta$,
while by virtue of the result \eqref{Finite temperature Interpolating: alpha beta L>2} the quantity $\Overline[2]{X}\Overline[2]{X}^\dagger$ turns out to be (at the first nontrivial order in the quark masses) $\Overline[2]{X}\Overline[2]{X}^\dagger \simeq \frac{F_X^2}{2}+ \frac{\kappa_1}{\lambda_X^2 F_X} \left(\frac{2B_m}{\sqrt{2}\lambda_\pi^2 B_\pi^2}\right)^L \det M \cos\theta$.
Putting all together, we see that the $\theta$ dependence of the minimum value of the potential is given, at the lowest order in the quark masses, by the following expression:
\begin{equation}\label{Finite temperature Interpolating: theta dependence of the potential L>2}
\begin{aligned}
V_{min}(\theta) &= \, const. - \frac{\kappa_1}{2\sqrt{2}}\left(\Overline[2]{X}^\dagger\det \Overline[2]{U} + \Overline[2]{X}\det \Overline[2]{U}^\dagger\right) + \ldots \\
&= \, const. -\frac{\kappa_1F_X}{2}
\left(\frac{2B_m}{\sqrt{2}\lambda_\pi^2 B_\pi^2}\right)^L
\det M\cos\theta + \ldots .
\end{aligned}
\end{equation}
From Eq. \eqref{Finite temperature Interpolating: theta dependence of the potential L>2} we can directly derive the following expressions for the topological susceptibility $\chi$ and the second cumulant $c_4$:
\begin{equation}\label{Finite temperature Interpolating: chi and c4 L>2}
\begin{aligned}
\chi = \left.\frac{\partial^2 V_{min}(\theta)}{\partial \theta^2}\right|_{\theta=0} &\simeq \frac{\kappa_1F_X}{2}\left(\frac{2B_m}{\sqrt{2}\lambda_\pi^2B_\pi^2}\right)^L \det M ,\\
c_4 = \left.\frac{\partial^4 V_{min}(\theta)}{\partial \theta^4}\right|_{\theta=0} &\simeq -\,\frac{\kappa_1F_X}{2}\left(\frac{2B_m}{\sqrt{2}\lambda_\pi^2B_\pi^2}\right)^L \det M .
\end{aligned}
\end{equation}
Comparing these last results with those that we have found in the $EL_\sigma$ model (for the case $L \ge 3$), we see that they coincide with each other (at least, at the leading order in the quark masses) provided that the parameter $\kappa$ in Eqs. \eqref{Finite temperature ELsm: theta dependence of potential L>2} and \eqref{Finite temperature ELsm: chi and c4 L>2} is identified with $\kappa_1 F_X/4$ (and is thus proportional to the $U(1)$ axial condensate).

\subsection{The special case $L=2$}

Being the quark-mass matrix ${\cal M} = M e^{i\frac{\theta}{2}}$ diagonal, and remembering what we have found for the nondiagonal elements of the matrix field $\Overline[2]{U}$ in the case of the $EL_\sigma$ model in Sec. 2.2, we can reasonably assume $\Overline[2]{U}$ to be diagonal since the beginning (i.e., $\bar{\delta}_1 = \bar{\delta}_2 = \bar{\pi}_1 = \bar{\pi}_2 = 0$). In other words, we take $\Overline[2]{U}$ and $\Overline[2]{X}$ in the form
\begin{equation}\label{Diagonal form of U L=2}
\Overline[2]{U}=\frac{1}{\sqrt{2}}\big[(\bar{\sigma}+i\bar{\eta})\mathbf{I} + (\bar{\delta}_3 + i \bar{\pi}_3)\,\tau_3\big] ,\quad \Overline[2]{X} =
\bar{\alpha} e^{i\bar{\beta}} .
\end{equation}
For what concerns the various terms of the potential in this case,
the only term which needs to be put in a new and more explicit form is the interaction term between $U$ and $X$, which turns out to be:
$\Overline[2]{X}^\dagger \det \Overline[2]{U} {+} \Overline[2]{X} \det \Overline[2]{U}^\dagger = \bar{\alpha}\left[(\bar{\sigma}^2 {-} \bar{\eta}^2 {-} \bar{\delta}_3^2 {+} \bar{\pi}_3^2)\cos\bar{\beta} {+} 2(\bar{\eta}\bar{\sigma} {-} \bar{\delta}_3\bar{\pi}_3)\sin\bar{\beta}\right]$.
Putting together all these results, we find the following expression for the potential:
\begin{equation}\label{Finite temperature Interpolating: explicit potential L=2}
\begin{aligned}
\bar{V} &= \frac{1}{8}\lambda_\pi^2 B_\pi^4 + \frac{1}{8}\Lambda_\pi^2(\bar{\sigma}^2 + \bar{\eta}^2 + \bar{\delta}_3^2 + \bar{\pi}_3^2)^2 + \frac{1}{2}\lambda_\pi^2(\bar{\sigma}^2\bar{\delta}_3^2 + 2\bar{\sigma}\bar{\eta}\bar{\delta}_3\bar{\pi}_3 + \bar{\eta}^2\bar{\pi}_3^2) \\
&+ \frac{1}{4}\lambda_\pi^2 B_\pi^2(\bar{\sigma}^2 + \bar{\eta}^2 + \bar{\delta}_3^2 + \bar{\pi}_3^2) + \frac{1}{4}\lambda_X^2\left(\bar{\alpha}^2-\frac{F_X^2}{2}\right)^2 +\frac{1}{2}A\bar{\beta}^2 \\
&- \frac{B_m}{2}\left[(m_u {+} m_d)\left(\bar{\sigma}\cos\frac{\theta}{2} {-} \bar{\eta}\sin\frac{\theta}{2}\right) {+} (m_u{-}m_d)\left(\bar{\delta}_3\cos\frac{\theta}{2} {-} \bar{\pi}_3\sin\frac{\theta}{2}\right)\right] \\
&- \frac{\kappa_1\bar{\alpha}}{2\sqrt{2}}\left[(\bar{\sigma}^2 - \bar{\eta}^2 - \bar{\delta}_3^2 + \bar{\pi}_3^2)\cos\bar{\beta} + 2(\bar{\eta}\bar{\sigma}-\bar{\delta}_3\bar{\pi}_3)\sin\bar{\beta}\right] .
\end{aligned}
\end{equation}
As usual, in order to find the minimum of the potential, we have to solve the following system of stationary-point equations:
\begin{equation}\label{Finite temperature Interpolating: minimization system L=2}
\left\{
\begin{aligned}
\left.\frac{\partial V}{\partial \sigma}\right|_S &= \frac{1}{2}\Lambda_\pi^2(\bar{\sigma}^2 + \bar{\eta}^2 + \bar{\delta}_3^2 + \bar{\pi}_3^2)\bar{\sigma} + \lambda_\pi^2(\bar{\sigma}\bar{\delta}_3^2 + \bar{\eta}\bar{\delta}_3\bar{\pi}_3) + \frac{1}{2}\lambda_\pi^2 B_\pi^2\bar{\sigma} \\
&-\frac{B_m}{2}(m_u + m_d)\cos\frac{\theta}{2}-\frac{\kappa_1\bar{\alpha}}{\sqrt{2}}(\bar{\sigma}\cos\bar{\beta}+ \bar{\eta}\sin\bar{\beta})=0 ,\\ \\
\left.\frac{\partial V}{\partial \eta}\right|_S &= \frac{1}{2}\Lambda_\pi^2(\bar{\sigma}^2 + \bar{\eta}^2 + \bar{\delta}_3^2 + \bar{\pi}_3^2)\bar{\eta} + \lambda_\pi^2(\bar{\sigma}\bar{\delta}_3\bar{\pi}_3 + \bar{\eta}\bar{\pi}_3^2) + \frac{1}{2}\lambda_\pi^2 B_\pi^2\bar{\eta} \\
&+\frac{B_m}{2}(m_u + m_d)\sin\frac{\theta}{2}-\frac{\kappa_1\bar{\alpha}}{\sqrt{2}}(-\bar{\eta}\cos\bar{\beta}+\bar{\sigma}\sin\bar{\beta})=0 ,\\ \\
\left.\frac{\partial V}{\partial \delta_3}\right|_S &= \frac{1}{2}\Lambda_\pi^2(\bar{\sigma}^2 + \bar{\eta}^2 + \bar{\delta}_3^2 + \bar{\pi}_3^2)\bar{\delta}_3 + \lambda_\pi^2(\bar{\sigma}^2\bar{\delta}_3 + \bar{\sigma}\bar{\eta}\bar{\pi}_3) + \frac{1}{2}\lambda_\pi^2 B_\pi^2\bar{\delta}_3 \\
&-\frac{B_m}{2}(m_u - m_d)\cos\frac{\theta}{2}-\frac{\kappa_1\bar{\alpha}}{\sqrt{2}}(-\bar{\delta}_3\cos\bar{\beta}-\bar{\pi}_3\sin\bar{\beta})=0 ,\\ \\
\left.\frac{\partial V}{\partial \pi_3}\right|_S &= \frac{1}{2}\Lambda_\pi^2(\bar{\sigma}^2 + \bar{\eta}^2 + \bar{\delta}_3^2 + \bar{\pi}_3^2)\bar{\pi}_3 + \lambda_\pi^2(\bar{\sigma}\bar{\eta}\bar{\delta}_3+\bar{\eta}^2\bar{\pi}_3) + \frac{1}{2}\lambda_\pi^2 B_\pi^2\bar{\pi}_3 \\
&+ \frac{B_m}{2}(m_u - m_d)\sin\frac{\theta}{2}-\frac{\kappa_1\bar{\alpha}}{\sqrt{2}}(\bar{\pi}_3\cos\bar{\beta}-\bar{\delta}_3\sin\bar{\beta})=0 ,\\ \\
\left.\frac{\partial V}{\partial \alpha}\right|_S &=\lambda_X^2\left(\bar{\alpha}^2-\frac{F_X^2}{2}\right)\bar{\alpha} \\
&- \frac{\kappa_1}{2\sqrt{2}}\left[(\bar{\sigma}^2 {-} \bar{\eta}^2 {-} \bar{\delta}_3^2 {+} \bar{\pi}_3^2)\cos\bar{\beta} + 2(\bar{\eta}\bar{\sigma}{-}\bar{\delta}_3\bar{\pi}_3)\sin\bar{\beta}\right]=0 ,\\ \\
\left.\frac{\partial V}{\partial \beta}\right|_S &= - \frac{\kappa_1\bar{\alpha}}{2\sqrt{2}}\left[-(\bar{\sigma}^2 {-} \bar{\eta}^2 {-} \bar{\delta}_3^2 {+} \bar{\pi}_3^2)\sin\bar{\beta} + 2(\bar{\eta}\bar{\sigma}{-}\bar{\delta}_3\bar{\pi}_3)\cos\bar{\beta}\right]+A\bar{\beta}=0 .
\end{aligned}
\right.
\end{equation}
Solving these equations at the first nontrivial order in the quark masses,
one finds that
\begin{equation}\label{Finite temperature Interpolating: solution of the fields L=2}
\begin{aligned}
\bar{\sigma} &\simeq \frac{B_m (m_u+m_d)}{\lambda_\pi^2 B_\pi^2 - \kappa_1 F_X}\cos\frac{\theta}{2} ,\qquad \bar{\eta} \simeq -\frac{B_m (m_u+m_d)}{\lambda_\pi^2 B_\pi^2 + \kappa_1 F_X}\sin\frac{\theta}{2} ,\\
\bar{\delta}_3 &\simeq \frac{B_m (m_u-m_d)}{\lambda_\pi^2 B_\pi^2 + \kappa_1 F_X}\cos\frac{\theta}{2} ,\qquad \bar{\pi}_3 \simeq -\frac{B_m (m_u-m_d)}{\lambda_\pi^2 B_\pi^2 - \kappa_1 F_X}\sin\frac{\theta}{2} ,\\
\bar{\alpha} &\simeq \frac{F_X}{\sqrt{2}} + \frac{\sqrt{2}\kappa_1^2\lambda_\pi^2 B_\pi^2}{\lambda_X^2 F_X (\lambda_\pi^4 B_\pi^4 - \kappa_1^2 F_X^2)^2}B_m^2(m_u^2+m_d^2) \\
&+ \frac{\sqrt{2}\kappa_1^2(\lambda_\pi^4 B_\pi^4+\kappa_1^2 F_X^2)}{\lambda_X^2 F_X^2 (\lambda_\pi^4 B_\pi^4 - \kappa_1^2 F_X^2)^2}B_m^2 m_u m_d \cos\theta ,\\
\bar{\beta} &\simeq -\frac{\kappa_1 F_X}{A}\frac{B_m^2 m_u m_d}{\lambda_\pi^4 B_\pi^4-\kappa_1^2 F_X^2}\sin\theta .
\end{aligned}
\end{equation}
Studying the matrix of the second derivatives of the potential with respect to the fields, one immediately verifies that this solution corresponds indeed to a minimum of the potential, provided that the condition $\lambda_\pi^2 B_\pi^2 > \kappa_1 F_X$, i.e., remembering Eq. \eqref{rho_pi for T>Tc}, ${\cal G}_\pi \equiv \kappa_1 F_X + 2\lambda_\pi^2 \rho_\pi < 0$, is satisfied.
As in the case of the $EL_\sigma$ model for $L=2$ (discussed in Sec. 2.2), the critical transition temperature $T_c$ is just defined by the condition ${\cal G}_\pi(T=T_c)=0$ and, assuming that $\kappa_1 F_X > 0$, this implies that (differently from the case $L \ge 3$) $T_c>T_{\rho_\pi}$ (see also Ref. \cite{MM2013} for a more detailed discussion on this question).

Substituting this solution into Eq. \eqref{Finite temperature Interpolating: explicit potential L=2}, and neglecting (for consistency) all the terms which are more than quadratic in the quark masses, we find the following $\theta$ dependence for the minimum value of the potential:
\begin{equation}\label{Finite temperature Interpolating: theta dependence of potential L=2}
\begin{aligned}
V_{min}(\theta) &= \frac{\lambda_\pi^2}{8}B_\pi^4 {+} \frac{\lambda_\pi^2}{4}B_\pi^2(\bar{\sigma}^2 {+} \bar{\eta}^2 {+} \bar{\delta}_3^2 {+} \bar{\pi}_3^2) {-}\frac{B_m}{2}\left[(m_u{+}m_d)\left(\bar{\sigma}\cos\frac{\theta}{2}{-} \bar{\eta}\sin\frac{\theta}{2}\right) \right. \\
&+ \left.(m_u {-} m_d)\left(\bar{\delta}_3\cos\frac{\theta}{2}{-} \bar{\pi}_3\sin\frac{\theta}{2}\right)\right] {-}\frac{\kappa_1 F_X}{4}(\bar{\sigma}^2 {-} \bar{\eta}^2 {-} \bar{\delta}_3^2 {+} \bar{\pi}_3^2) + O(m^3) \\
&= \, const. -\frac{\kappa_1 F_X B_m^2 m_u m_d}{\lambda_\pi^4 B_\pi^4 - \kappa_1^2 F_X^2}\cos\theta + O(m^3) .
\end{aligned}
\end{equation}
Also in this case, we notice that this potential, as well as the expressions \eqref{Finite temperature Interpolating: solution of the fields L=2} for $\bar{\sigma}$, $\bar{\eta}$, $\bar{\delta}_3$, and $\bar{\pi}_3$, coincide exactly (at least, at the leading order in the quark masses) with the corresponding expressions \eqref{Finite temperature ELsm: solution of the fields L=2} and \eqref{Finite temperature ELsm: theta dependence of potential L=2} that we have found in the $EL_{\sigma}$ model, provided that the constant $\kappa$ is identified with $\kappa_1 F_X/4$ (and is thus proportional to the $U(1)$ axial condensate). The same consideration also applies, of course, to the results for the topological susceptibility and for the second cumulant:
\begin{equation}\label{Finite temperature Interpolating: chi and c4 L=2}
\begin{aligned}
\chi = \left.\frac{\partial^2 V_{min}(\theta)}{\partial \theta^2}\right|_{\theta=0} &\simeq \frac{\kappa_1 F_X B_m^2}{\lambda_\pi^4 B_\pi^4 - \kappa_1^2 F_X^2} m_u m_d ,\\
c_4 = \left.\frac{\partial^4 V_{min}(\theta)}{\partial \theta^4}\right|_{\theta=0} &\simeq -\frac{\kappa_1 F_X B_m^2}{\lambda_\pi^4 B_\pi^4 - \kappa_1^2 F_X^2} m_u m_d .
\end{aligned}
\end{equation}

\section{Conclusions: summary and analysis of the results}

In this conclusive section we summarize and critically comment on the results that we have found, indicating also some possible future perspectives.

Two basic remarks must be made about our results.
First (as already observed at the end of Secs. 3.1 and 3.2), the results that we have found (both in the case $L \ge 3$ and in the case $L=2$) for the vacuum energy density $\epsilon_{vac}(\theta) = V_{min}(\theta)$ (and, as a consequence, for the topological susceptibility $\chi$ and the second cumulant $c_4$) in the $EL_\sigma$ model and in the interpolating model are exactly the same, provided that the parameter $\kappa$ in Eqs. \eqref{Finite temperature ELsm: theta dependence of potential L>2}--\eqref{Finite temperature ELsm: chi and c4 L>2} and \eqref{Finite temperature ELsm: theta dependence of potential L=2}--\eqref{Finite temperature ELsm: chi and c4 L=2} is identified with $\kappa_1 F_X/4$ (and is, therefore, proportional to the $U(1)$ axial condensate).
In fact, we have found that
\begin{equation}\label{theta dependence of epsilon}
\epsilon_{vac}(\theta) \simeq const. - K \cos\theta ,
\end{equation}
and, therefore,
\begin{equation}
\chi = \left.\frac{\partial^2 \epsilon_{vac}(\theta)}{\partial \theta^2}\right|_{\theta=0} \simeq K ,\qquad
c_4 = \left.\frac{\partial^4 \epsilon_{vac}(\theta)}{\partial \theta^4}\right|_{\theta=0} \simeq -K ,
\end{equation}
where, for $L \ge 3$,
\begin{equation}\label{K for L>2}
K_{(L \ge 3)} = 2\kappa
\left( \frac{2 B_m}{\sqrt{2} \lambda_\pi^2 B_\pi^2} \right)^L \det M
= \frac{\kappa_1 F_X}{2}
\left( \frac{2 B_m}{\sqrt{2} \lambda_\pi^2 B_\pi^2} \right)^L \det M ,
\end{equation}
and, for $L=2$,
\begin{equation}\label{K for L=2}
K_{(L=2)} = \frac{4 \kappa B_m^2}{\lambda_\pi^4 B_\pi^4 - 16\kappa^2} m_u m_d
= \frac{\kappa_1 F_X B_m^2}{\lambda_\pi^4 B_\pi^4 - \kappa_1^2 F_X^2} m_u m_d .
\end{equation}
This result is, of course, in agreement with what we have already observed in the Introduction [see, in particular, Eq. \eqref{IM to ELsm}],
but we want to emphasize that it is even \emph{stronger} than the correspondence \eqref{IM to ELsm}, since it is valid \emph{regardless} of the parameters $\lambda_X$ and $A$ of the interpolating model (which do \emph{not} appear in the above-written expressions for $\epsilon_{vac}(\theta)$, $\chi$, and $c_4$).
Taking into account also the results that were found in Ref. \cite{LM2018}, we now clearly see that the so-called ``interpolating model'' indeed approximately ``interpolates'' between the WDV model at $T=0$ (for $\omega_1=1$ it reproduces the same expressions for $\chi$ and $c_4$ of the WDV model) and the $EL_\sigma$ model at $T > T_c$ (where $\omega_1=0$).

We also observe that the result \eqref{K for L=2}, for the special case $L=2$, can be rewritten in the following more interesting and enlightening way:
\begin{equation}\label{K for L=2 bis}
K_{(L=2)} \simeq \frac{M_\eta^2 - M_\sigma^2}{4 M_\eta^2 M_\sigma^2} (B_m m_u) (B_m m_d) ,
\end{equation}
in terms of the masses of the scalar and pseudoscalar mesonic excitations, which, at the leading order in the quark masses, are given by \cite{MM2013}
\begin{equation}
\begin{aligned}
M_\sigma^2 = M_\pi^2 &\simeq \frac{1}{2} (\lambda_\pi^2 B_\pi^2 - 4\kappa) = \frac{1}{2} (\lambda_\pi^2 B_\pi^2 - \kappa_1 F_X) ,\\
M_\eta^2 = M_\delta^2 &\simeq \frac{1}{2} (\lambda_\pi^2 B_\pi^2 + 4\kappa) = \frac{1}{2} (\lambda_\pi^2 B_\pi^2 + \kappa_1 F_X) .
\end{aligned}
\end{equation}
The second important remark that we want to make about our results is that both the $\theta$ dependence of $\epsilon_{vac}(\theta)$ in Eq. \eqref{theta dependence of epsilon} and the quark-mass dependence of the coefficient $K$ (proportional to $\det M$) are in agreement with the corresponding results found using the so-called ``dilute instanton-gas approximation'' (DIGA) \cite{GPY1981}.
Of course, we cannot make any more quantitative statements about the comparison of our value of $K$ with the corresponding value $K_{inst}$ in DIGA, or about its dependence on the temperature $T$.\\
In this respect, recent lattice investigations have shown contrasting results.
Some studies have shown a considerable agreement with the DIGA prediction even in the region right above $T_c$ \cite{lattice_1,lattice_2} or in the region above $1.5 T_c$ \cite{lattice_3}, while other studies \cite{lattice_4,lattice_5} have found appreciable deviations from the DIGA prediction for temperatures $T$ up to two or three times $T_c$.
The situation is thus controversial and calls for further and more accurate studies (in this respect, see also Ref. \cite{DDSX2017}).

Concerning, instead, the limits of validity of our analytical results \eqref{theta dependence of epsilon}--\eqref{K for L=2}, we recall that they were obtained at the first nontrivial order in an expansion in the quark masses. Therefore, both the coincidence between the results in the two models and the agreement with the $\theta$ dependence predicted by DIGA are valid in this approximation and it would be interesting to investigate how strongly these results are modified going beyond the leading order in the quark masses.
(It is reasonable to suspect that this approximation makes sense for $T - T_c \gg m_f$, but \emph{not} for $T$ close to $T_c$, i.e., for $T - T_c \lesssim m_f$.)\\
A complete and detailed study of the $\theta$ dependence of $\epsilon_{vac}(\theta)$ both for the $EL_\sigma$ model and the interpolating model, not limited to the leading order in the quark masses, is beyond the scope of the present paper and is left for future works.\\
A first step in this direction has been, however, already done in the Appendix of the present paper, where an ``exact'' expression for the topological susceptibility $\chi$ for $T>T_c$ has been derived, both for the interpolating model (considering the effective Lagrangian in the form \eqref{Interpolating model Lagrangian with Q}--\eqref{potential of the interpolating model}, where the field variable $Q(x)$ has not yet been integrated out) and, making use of the correspondence \eqref{IM to ELsm}, also for the $EL_\sigma$ model.
The expressions of $\chi$ for the two models are reported in Eqs. \eqref{chi_IM} and \eqref{chi_ELsm} respectively. We see that they are slightly different, but if one expands at the leading order in the quark masses, one easily verifies that they both tend to the same limit (with the identification $\kappa = \kappa_1 F_X/4$), given by Eqs. \eqref{Finite temperature ELsm: chi and c4 L>2} and \eqref{Finite temperature Interpolating: chi and c4 L>2} for $L \ge 3$, and by Eqs. \eqref{Finite temperature ELsm: chi and c4 L=2} and \eqref{Finite temperature Interpolating: chi and c4 L=2} for $L=2$.\\
A necessary condition for this approximation to be valid is, of course, that [see Eqs. \eqref{chi_IM}--\eqref{det Lambda and det S}] $\det {\cal S} \ll \frac{A}{\bar{\alpha}^2} \det \Lambda$, which, by virtue of Eq. \eqref{chi_IM}, implies $\chi \ll A$.\\
(In the opposite extreme case, if we formally let $A \to 0$, keeping all the rest fixed, we would obtain that $\chi \simeq A \to 0$.)\\
While at $T=0$ this condition is reasonably satisfied, since in that case one identifies $A$ with the \emph{pure-gauge} topological susceptibility and [see Ref. \cite{LM2018} and references therein] $\chi(T=0) \simeq (75~{\rm MeV})^4$, $A(T=0) \simeq (180~{\rm MeV})^4$, its validity at finite temperature, above $T_c$, is, instead, questionable.
(For example, it is not even clear if, in our phenomenological Lagrangian for the interpolating model at finite temperature, the parameter $A(T)$ can be simply identified with the \emph{pure-gauge} topological susceptibility.)\\
We hope that future works (both analytical and numerical) will be able to shed light on these questions.

\section*{Acknowledgments}

The author is extremely grateful to Francesco Luciano for his help during the
initial stage of the work.

\newpage

\renewcommand{\thesection}{}
\renewcommand{\thesubsection}{A.\arabic{subsection} }
 
\pagebreak[3]
\setcounter{section}{1}
\setcounter{equation}{0}
\setcounter{subsection}{0}
\setcounter{footnote}{0}

\begin{flushleft}
{\Large\bf \thesection Appendix A: ``Exact'' expression for the topological susceptibility above $T_c$}
\end{flushleft}

\renewcommand{\thesection}{A}

\noindent
In the interpolating model, it is possible to derive the two--point function of $Q(x)$ (i.e., the topological susceptibility $\chi$) at $\theta=0$ in another (and even more direct) way, considering the effective Lagrangian in the form \eqref{Interpolating model Lagrangian with Q}--\eqref{potential of the interpolating model}, where the field variable $Q(x)$ has not yet been integrated out (and $\theta$ is fixed to be equal to zero, by putting $\mathcal{M} = M$). Clearly,
\begin{equation}\label{chi(k)}
\chi(k)\equiv-i\int d^{4}x\;e^{ikx}\langle{TQ(x)Q(0)}\rangle=
({\cal K}^{-1}(k))_{Q,Q},
\end{equation}
where ${\cal K}^{-1}(k)$ is the inverse of the matrix ${\cal K}(k)$ associated
with the quadratic part of the Lagrangian \eqref{Interpolating model Lagrangian with Q}--\eqref{potential of the interpolating model} in the momentum space, for the ensemble of pseudoscalar fields
$(Q,~S_X,~b_{11},~b_{12},\ldots)$ [see Eqs. \eqref{linear parametrization} and \eqref{Parametrization of the field X}, with $\alpha \equiv \bar{\alpha} + h_X$ and $\beta \equiv S_X/\bar{\alpha}$; the contribution of the scalar fields $(h_X,~a_{11},~a_{12},\ldots)$ is block diagonal and, therefore, can be trivially factorized out]:
\begin{equation}\label{matrix K(k)}
\mathcal{K}(k)=
\begin{pmatrix}
\frac{1}{A} & -\frac{1}{\bar{\alpha}} & 0 & \ldots \\
-\frac{1}{\bar{\alpha}} & {\cal R}(k)_{X,X} & {\cal R}(k)_{X,11} & \ldots \\
0 & {\cal R}(k)_{11,X} & {\cal R}(k)_{11,11} & \ldots \\
\vdots & \vdots & \vdots & \ddots \\
\end{pmatrix} ,
\end{equation}
where
\begin{equation}\label{matrix R(k)}
{\cal R}(k)= k^2 {\bf I} - {\cal S}, \quad {\cal S} =
\begin{pmatrix}
m^2_0 & \mathcal{O}(m^{L-1}) & \ldots \\
\mathcal{O}(m^{L-1}) & \Lambda_{11,11} & \ldots \\
\vdots & \vdots & \ddots \\
\end{pmatrix} ,
\end{equation}
and (assuming that $L \ge 3$)
\begin{equation}\label{m_0^2 and Lambda}
m_0^2 \equiv \frac{\kappa_1}{\sqrt{2} \bar{\alpha}} \det \Overline[2]{U} ,\quad
\Lambda_{ij,lm} = \frac{1}{2} \lambda_\pi^2 B_\pi^2 \delta_{il} \delta_{jm}
+ \ldots ,
\end{equation}
with
\begin{equation}\label{U_bar and alpha_bar}
\Overline[2]{U} = \frac{2 B_m}{\sqrt{2} \lambda_\pi^2 B_\pi^2} M + \ldots ,\quad
\bar{\alpha} = \frac{F_X}{\sqrt{2}} + \mathcal{O}(\det M) .
\end{equation}
Performing explicitly the computation, one finds that
\begin{equation}
\chi(k) = ({\cal K}^{-1}(k))_{Q,Q}
= \frac{\det {\cal R}(k)}{\det \mathcal{K}(k)}
= A \frac{\det \mathcal{R}(k)}{\det \tilde{\cal R}(k)} ,
\end{equation}
having defined
\begin{equation}\label{matrix Rtilde(k)}
\tilde{\cal R}(k)= k^2 {\bf I} - \tilde{\cal S}, \quad \tilde{\cal S} =
\begin{pmatrix}
m_0^2 + \frac{A}{\bar{\alpha}^2} & \mathcal{O}(m^{L-1}) & \ldots \\
\mathcal{O}(m^{L-1}) & \Lambda_{11,11} & \ldots \\
\vdots & \vdots & \ddots \\
\end{pmatrix} ,
\end{equation}
so that $\det \tilde{\cal R}(k) = \det {\cal R}(k) - \frac{A}{\bar{\alpha}^2}
\det (k^2 {\bf I} - \Lambda)$.
In particular, putting $k=0$, the following expression for the topological
susceptibility is found:
\begin{equation}\label{chi_IM}
\chi \equiv \chi(k=0)
= A \frac{\det {\cal S}}{\det \tilde{\cal S}}
= A \frac{\det {\cal S}}{\det {\cal S} + \frac{A}{\bar{\alpha}^2} \det \Lambda} .
\end{equation}
This expression has been obtained for the interpolating model, but, as
explained in the Introduction [see, in particular, Eq. \eqref{IM to ELsm}],
if we take the formal limits $\lambda_X \to \infty$ and $A \to \infty$
(having already fixed $\omega_1=0$, since we are at $T>T_c$),
we also obtain the expression for the topological susceptibility in the $EL_\sigma$ model:
\begin{equation}\label{chi_ELsm}
\chi_{(EL_\sigma)} = \frac{\bar{\alpha}^2 \det {\cal S}}{\det \Lambda}
= \frac{\det \Overline[2]{\cal S}}{\det \Lambda} ,
\end{equation}
where now $\bar{\alpha} = \frac{F_X}{\sqrt{2}}$ and
\begin{equation}
\Overline[2]{\cal S} =
\begin{pmatrix}
\Overline[2]{m}_0^2 & \bar{\alpha}\mathcal{O}(m^{L-1}) & \ldots \\
\bar{\alpha}\mathcal{O}(m^{L-1}) & \Lambda_{11,11} & \ldots \\
\vdots & \vdots & \ddots \\
\end{pmatrix} ,
\end{equation}
with
\begin{equation}
\Overline[2]{m}_0^2 \equiv \bar{\alpha}^2 m_0^2 =
\frac{\kappa_1 \bar{\alpha}}{\sqrt{2}} \det \Overline[2]{U}
= 2 \kappa \det \Overline[2]{U} ,
\end{equation}
having identified, as usual, $\kappa \equiv \frac{\kappa_1 \bar{\alpha}}{2 \sqrt{2}} = \frac{\kappa_1 F_X}{4}$.\\
Even with this identification, the two expressions \eqref{chi_IM} and
\eqref{chi_ELsm} are slightly different, but if one expands at the leading
order in the quark masses, using the fact that [see Eqs.
\eqref{m_0^2 and Lambda} and \eqref{U_bar and alpha_bar}]:
\begin{equation}\label{det Lambda and det S}
\det \Lambda = \left( \frac{1}{2} \lambda_\pi^2 B_\pi^2 \right)^{L^2} + \ldots,
\quad \det {\cal S} = \frac{\kappa_1}{F_X}
\left( \frac{2 B_m}{\sqrt{2} \lambda_\pi^2 B_\pi^2} \right)^L
 \left( \frac{1}{2} \lambda_\pi^2 B_\pi^2 \right)^{L^2} \det M + \ldots,
\end{equation}
one easily verifies that they both tend to the same limit,
given by Eqs. \eqref{Finite temperature ELsm: chi and c4 L>2}
and \eqref{Finite temperature Interpolating: chi and c4 L>2}:\footnote{This approximate
expression [but not the ``exact'' expressions \eqref{chi_IM} and \eqref{chi_ELsm}]
was derived (for the interpolating model) also in Ref. \cite{EM1994}.}
\begin{equation}
\chi \simeq \frac{\kappa_1 F_X}{2}
\left( \frac{2 B_m}{\sqrt{2} \lambda_\pi^2 B_\pi^2} \right)^L \det M
= 2\kappa
\left( \frac{2 B_m}{\sqrt{2} \lambda_\pi^2 B_\pi^2} \right)^L \det M .
\end{equation}
Similar results come out also in the special case $L=2$.
In particular, the expressions \eqref{chi_IM} and \eqref{chi_ELsm} (for the
topological susceptibility in the interpolating model and in the $EL_\sigma$
model, respectively) are valid also in this case, provided that one uses
for the matrices ${\cal S}$, $\Overline[2]{\cal S}$, and $\Lambda$ the following
expressions [referring to the ensemble of pseudoscalar fields
$(S_X,~\eta,~\pi_3)$: see Ref. \cite{MM2013} for further details]:
\begin{equation}\label{matrix S L=2}
{\cal S}_{(L=2)} = \left(
\begin{matrix}
m_0^2 & -\frac{\kappa_1}{\sqrt{2}}\bar{\sigma} &
\frac{\kappa_1}{\sqrt{2}}\bar{\delta}_3 \\
-\frac{\kappa_1}{\sqrt{2}}\bar{\sigma} & \Lambda_{11} & \Lambda_{12} \\ 
\frac{\kappa_1}{\sqrt{2}}\bar{\delta}_3 & \Lambda_{21} & \Lambda_{22}
\end{matrix}
\right) ,\qquad
\Overline[2]{\cal S}_{(L=2)} = \left(
\begin{matrix}
\Overline[2]{m}_0^2 & -2\kappa\bar{\sigma} & 2\kappa\bar{\delta}_3 \\
-2\kappa\bar{\sigma} & \Lambda_{11} & \Lambda_{12} \\
2\kappa\bar{\delta}_3 & \Lambda_{21} & \Lambda_{22}
\end{matrix}
\right) ,
\end{equation}
and
\begin{equation}\label{matrix Lambda L=2}
\Lambda_{(L=2)} = \left(
\begin{matrix}
\frac{1}{2}(\lpq\bpq+\kappa_1\sqrt{2}\bar{\alpha}) + \Delta & \lpq\bar{\delta}_3\bar{\sigma} \\
\lpq\bar{\delta}_3\bar{\sigma} & \frac{1}{2}(\lpq\bpq-\kappa_1\sqrt{2}\bar{\alpha}) + \Delta
\end{matrix}
\right) ,
\end{equation}
where $\Delta \equiv \frac{1}{2}\Lpq(\bar{\sigma}^2+\bar{\delta}_3^2)$
(having defined $\Lpq \equiv \lpq + 2\lambda_\pi^{'2}$) and
\begin{equation}
m_0^2 \equiv \frac{\kappa_1}{2\sqrt{2}\bar{\alpha}}(\bar{\sigma}^2-\bar{\delta}_3^2) ,\qquad
\Overline[2]{m}_0^2 \equiv \kappa(\bar{\sigma}^2-\bar{\delta}_3^2) ,
\end{equation}
with
\begin{equation}
\bar{\sigma} = \frac{B_m (m_u+m_d)}{\lpq\bpq - \kappa_1 F_X} + \ldots ,\quad
\bar{\delta}_3 = \frac{B_m (m_u-m_d)}{\lpq\bpq + \kappa_1 F_X} + \ldots ,\quad
\bar{\alpha} = \frac{F_X}{\sqrt{2}} + \mathcal{O}(m^2) .
\end{equation}
Using these expressions, one finds that, at the leading order in the quarks masses,
\begin{equation}
\det \Lambda = \frac{1}{4} (\lambda_\pi^4 B_\pi^4 - \kappa_1^2 F_X^2) +
\mathcal{O}(m^2) ,\qquad
\det {\cal S} = \frac{\kappa_1 B_m^2}{2 F_X} m_u m_d + \mathcal{O}(m^3) ,
\end{equation}
so that, also in the special case $L=2$, one easily verifies that the two
expressions \eqref{chi_IM} and \eqref{chi_ELsm} tend to the same limit,
given by Eqs. \eqref{Finite temperature ELsm: chi and c4 L=2}
and \eqref{Finite temperature Interpolating: chi and c4 L=2}:
\begin{equation}
\chi \simeq
\frac{\kappa_1 F_X B_m^2}{\lambda_\pi^4 B_\pi^4 - \kappa_1^2 F_X^2} m_u m_d
= \frac{4 \kappa B_m^2}{\lambda_\pi^4 B_\pi^4 - 16\kappa^2} m_u m_d .
\end{equation}

\newpage

\renewcommand{\Large}{\large}

\end{document}